\documentclass[a4paper,11pt]{article}

\usepackage[a4paper]{geometry}
\usepackage{color}
\usepackage{graphicx}
\usepackage{amsmath}
\usepackage{amsfonts}
\usepackage{amsthm}
\usepackage{amscd}
\usepackage{epsfig}
\usepackage{amssymb}
\usepackage{slashed}
\usepackage{tabularx}
\usepackage{calligra}
\usepackage{enumerate}
\usepackage{hyperref}
\usepackage{float}
\usepackage{stackrel}
\usepackage[T1]{fontenc}
\usepackage{longtable}
\usepackage{amsthm}
\usepackage{mathrsfs}
\usepackage{verbatim} % for comments
\usepackage{fancyhdr}
\usepackage{tikz}
\usepackage{tikz-cd}
\usepackage{slashed}
\usetikzlibrary{matrix}
\usetikzlibrary{positioning}
\usepackage[all]{xy}
\usepackage{units}
\usepackage{textcomp}
\usepackage{turnstile}
\usepackage{vmargin}
\usepackage{anysize}
\usepackage{ stmaryrd }
\usepackage{tikz-3dplot}

\setmargins{2,5cm}       % left margin
{1,5cm}                        %  top margin
{16cm}                      %  text widht
{23,42cm}                    % text height
{10pt}                           % header height
{1cm}                           % space between text and header
{0pt}                             % footer height
{2cm}                           %  space between text and footer

\theoremstyle{plain}

\newtheorem{proposition}{Proposition}

\theoremstyle{definition}
\newtheorem*{proposition*}{Proposition}

\def\d{{\rm d}}
\def\i{{\rm i}}

\def\CM{\mathbb{CM}}
\def\CP{\mathbb{CP}}
\def\PT{\mathbb{PT}}

\def\rc{r_{\rm c}}
\def\O{\mathcal{O}}

\begin{document}

\title{\bf Complex conformal transformations and zero-rest-mass fields}

\author{Bernardo Araneda\footnote{Email: \texttt{bernardo.araneda@aei.mpg.de}}  
\\
Max-Planck-Institut f\"ur Gravitationsphysik \\ 
(Albert-Einstein-Institut), Am M\"uhlenberg 1, \\
D-14476 Potsdam, Germany}

\date{\today}

\maketitle

\begin{abstract} 
We give a simple prescription for relating different solutions to the zero-rest-mass field equations in conformally flat space-time via complex conformal transformations and changes in reality conditions. We give several examples including linearized black holes. In particular, we show that the linearized Pleba\'nski-Demia\'nski and Schwarzschild fields are related by a complex translation and a complex special conformal transformation. Similar results hold for the linearized Kerr and C-metric fields, and for a peculiar toroidal singularity.
\end{abstract}

\section{Introduction}

The Newman-Janis complex shift \cite{NewmanJanis} 
is a method for obtaining the Kerr solution 
to the Einstein vacuum equations from the Schwarzschild solution 
via a complex coordinate transformation. 
The apparent arbitrariness in the way in which some of the metric functions must be complexified makes it difficult to establish whether it has a deep geometric origin \cite{Flaherty}.
The linear version of it, however, 
can be understood as a simple complex translation 
$z \to z-\i a$ \cite{Newman2002}. 
The interest in this shift has been recently renewed in view of 
its applications to scattering amplitudes 
\cite{ArkaniHamed, Guevaraetal, Buonanno}.

In this note we show that a complex translation followed by a 
complex special conformal transformation applied to the 
linearized Schwarzschild field produces the linearized 
Pleba\'nski-Demia\'nski field (which is the linear limit of the most general type D vacuum space-time), and we furthermore show that this is 
just an example of a general and simple procedure in 
twistor space that applies to generic zero-rest-mass fields. 
This allows us to uncover complex coordinate transformations 
between, for example, the (linearized) Kerr and C-metric fields and a curious toroidal structure, 
as well as complex transformations applied to constant fields, 
hopfions/knotted fields, plane waves, etc., with arbitrary spin 
and algebraic type, as part of a unified framework. 

It is interesting to note that the basic idea in the Newman-Janis 
shift can be traced back to at least 1887, when Appell \cite{Appell} noticed that the point singularity $\{x=y=z=0\}$ of 
the fundamental solution $(x^2+y^2+z^2)^{-1/2}$ to the Laplace 
equation in $\mathbb{R}^3$ is mapped to a ring singularity 
$\{x^2+y^2=a^2, \ z=0\}$ under the complex translation $z\to z-\i a$.
Synge \cite{Synge} generalized this to remove the light-cone 
singularity of the fundamental solution $(t^2-x^2-y^2-z^2)^{-1}$ 
to the relativistic wave equation in Minkowski space-time.
Complex translations, and, more generally, complex Poincar\'e transformations, were then recognized by Trautman \cite{Trautman} 
as a powerful tool for the generation of new solutions to the scalar, Maxwell, and linearized gravity field equations in flat space-time; 
see also \cite{Newman}.

\section{Preliminaries}
\label{Sec:Preliminaries}

We will use twistor methods. For background on the aspects of 
twistor theory relevant to this work, we refer to 
\cite{PMC73, PR2, HuggettTod, Adamo}.
We include appendix \ref{Appendix} with some spinor conventions.

Let $\CM$ be complexified Minkowski space-time, with 
flat holomorphic metric $\eta=\d t^2 - \d x^2 - \d y^2 - \d z^2$. 
The twistor space of $\CM$ is $\PT=\CP^3\backslash\CP^1$, 
with homogeneous coordinates $Z^{\alpha}=(Z^0,Z^1,Z^2,Z^3)$. 
In 2-spinor notation, space-time coordinates are 
encoded in a $2\times2$ matrix $x^{AA'}$, and points of $\PT$
are represented by $Z^{\alpha}=(\omega^{A},\pi_{A'})$, with 
$Z^0=\omega^0$, $Z^1=\omega^1$, $Z^2=\pi_{0'}$, $Z^3=\pi_{1'}$.
The two spaces are related by the incidence relation 
\begin{align}
\omega^{A}=\i x^{AA'}\pi_{A'}. \label{IR}
\end{align}
The $\CP^1$ removed in the definition  
$\PT=\CP^3\backslash\CP^1$ corresponds to the set
$\{Z^2=Z^3=0\}$ (i.e. $\{\pi_{A'}=0\}$). 
This gives a fibration $\PT\to\CP^1$, where 
$\pi_{A'}$ are inhomogeneous coordinates on the base, and 
$\omega^A$ are coordinates on the fibers.

From \eqref{IR} one deduces that a point $x^{AA'}\in\CM$ 
corresponds to a holomorphic linear Riemann sphere 
$L_{x}=\CP^1\subset\PT$ (a twistor line), while a point 
$Z^{\alpha}\in\PT$ corresponds to a totally null 2-surface 
in $\CM$ (an $\alpha$-surface). 
The set $\{Z^2=Z^3=0\}$ removed from $\PT$ is 
a twistor line ${\bf I}$ in the twistor space of conformally 
compactified Minkowski space-time $\CM^{\sharp}$. 
This twistor space is the compactification $\CP^3$, the
line ${\bf I}$ corresponds to the vertex $I$ of the null cone at infinity 
and it can also be represented by the infinity twistor 
$I_{\alpha\beta}\d{Z}^{\alpha}\wedge\d{Z}^{\beta}=
2 \: \d{Z}^{2}\wedge\d{Z}^{3}$.

Two twistor lines $L_x,L_y$ intersect if and only if the associated space-time points $x,y$ are null-separated. This means that the conformal structure of space-time is 
encoded in the intersection of twistor lines in $\PT$.
More generally, twistor theory is conformally invariant, and twistors 
can be understood as the spinors of the (complexified) 
conformal group ${\rm SL}(4,\mathbb{C})$. 
In other words, twistor space carries a 
representation of ${\rm SL}(4,\mathbb{C})$: 
a complex {\em linear} transformation 
\begin{align}
Z^{\alpha} \mapsto T^{\alpha}{}_{\beta}Z^{\beta}, 
\quad T^{\alpha}{}_{\beta}\in{\rm SL}(4,\mathbb{C}),
\label{LinearT}
\end{align}
corresponds to a complex {\em conformal} transformation on space-time \cite{Penrose1967}
(that is, to an element of the 15-complex-dimensional group of complex Poincar\'e transformations, complex dilations, and 
complex special conformal transformations). 
More precisely, the conformal group acts on the compactified 
space $\CM^{\sharp}$, since conformal inversions interchange 
the origin with $I$.
The subgroup of ${\rm SL}(4,\mathbb{C})$ that leaves the line 
${\bf I}$ in $\CP^3$ invariant is the Poincar\'e group.

We can express any $T^{\alpha}{}_{\beta}$ as a matrix 
\begin{align}
 T^{\alpha}{}_{\beta} = 
 \left(\begin{matrix} \theta^{A}{}_{B} & \tau^{AB'} \\
 \nu_{A'B} & \tilde{\theta}_{A'}{}^{B'} \end{matrix} \right).
 \label{genericT}
\end{align}
To describe the action on space-time coordinates, we separate into 
three cases: 
$(i)$ $\tau^{AB'}=0=\nu_{A'B}$, 
$(ii)$ $\theta^{A}{}_{B}=\delta^{A}_{B}$, 
$\tilde\theta_{A'}{}^{B'}=\delta_{A'}^{B'}$, $\nu_{AB'}=0$, and 
$(iii)$ $\theta^{A}{}_{B}=\delta^{A}_{B}$, 
$\tilde\theta_{A'}{}^{B'}=\delta_{A'}^{B'}$, $\tau^{AB'}=0$. 
Then we find:
\begin{subequations}\label{Tx}
\begin{align}
(i) \qquad x'^{AA'} ={}& \theta^{A}{}_{B}x^{BB'}(\tilde\theta^{-1})_{B'}{}^{A'}, \\
(ii) \qquad x'^{AA'} ={}& x^{AA'}+\xi^{AA'}, \qquad \xi^{AA'}=-\i\tau^{AA'}, \\
(iii) \qquad x'^{AA'} ={}& \frac{x^{AA'}-(x_bx^b)s^{AA'}}
 {(x_bx^b)(s_cs^c)-2x_bs^b+1}, \qquad 
 s^{AA'} = -\tfrac{\i}{2}\nu^{AA'}
\end{align}
\end{subequations}
Thus, we see that $\theta^{A}{}_{B}$ and $\tilde\theta_{A'}{}^{B'}$ describe left and right Lorentz transformations and dilations, 
$\tau^{AB'}$ corresponds to translations and $\nu_{AB'}$ to special conformal transformations. 

\smallskip
Let $h>0$ be a positive half-integer number.
A zero-rest-mass field of helicity (or spin) $h$ is a totally symmetric 
spinor $\varphi_{A'...K'}$ with $2h$ indices such that
\begin{align}
\nabla^{AA'}\varphi_{A'...K'}=0. \label{ZRMF}
\end{align}
For $h=0$, the corresponding equation is $\Box\varphi=0$. 
The cases $h=\frac{1}{2},1,\frac{3}{2},2$ describe (massless) Dirac, Maxwell, Rarita-Schwinger,
and linearized gravitational fields, respectively. 
The field equations \eqref{ZRMF} are conformally invariant, 
as long as $\varphi_{A'...K'}$ has conformal weight $-1$.
A classical result from twistor theory (see e.g. \cite{HuggettTod})  establishes that the set of 
zero-rest-mass fields is isomorphic to the \v{C}ech 
cohomology group 
$\breve{H}^{1}(\PT,\O(-2h-2))$, where $\O(k)$ 
is the sheaf of holomorphic functions on $\PT$ that are homogeneous of 
degree $k$. An explicit representation of this isomorphism is the 
Penrose transform: if $x^{AA'}\in\CM$ and $L_x=\CP^1$ is the associated 
twistor line, then any solution to \eqref{ZRMF} can be written as 
\begin{align}
\varphi_{A'...K'}(x) = \frac{1}{2\pi\i}\oint_{\Gamma}f(Z)\big|_{L_{x}}
\pi_{A'}...\pi_{K'}\pi_{L'}\d\pi^{L'},
\label{Penrosetransform}
\end{align}
where the twistor function $f$ is homogeneous of 
degree $-2h-2$ and holomorphic except for a certain singularity 
region, and the contour $\Gamma\subset L_{x}$ is such that 
it surrounds the singularities of $f$, see \cite[Section 6.10]{PR2}.
The relationship between $\varphi_{A'...K'}$ and $f$ is not unique, but $f$ is just a representative of a cohomology class in 
$\breve{H}^{1}(\PT,\O(-2h-2))$. 
For practical calculations, however, one can
work with representatives, as long as the 
corresponding space-time result is cohomological invariant.
This will be the case for the applications considered in this work.

\subsubsection*{Main idea}

The basic observation in this work is the following. 
Consider a twistor function $f$, which generates a zero-rest-mass 
field $\varphi_{A'...K'}$ via the Penrose transform. 
Consider also a linear transformation \eqref{LinearT} in $\PT$, 
and put $Z'^{\alpha}=T^{\alpha}{}_{\beta}Z^{\beta}$.
Define 
\begin{align}
f'(Z):=f(Z'). \label{BasicTr}
\end{align}
The Penrose transform of $f'$ will generate a new zero-rest-mass 
field $\varphi'_{A'...K'}$ (of the same helicity).
But we mentioned that \eqref{LinearT} corresponds to a 
complex conformal transformation on space-time.
Imposing different reality conditions on space-time coordinates 
before and after the transformation, this means that the 
fields $\varphi_{A'...K'}$ and $\varphi'_{A'...K'}$ 
will be two different solutions, that can be 
mapped to each other via a complex conformal transformation.
We will illustrate this with several examples in the next section.
We note that, even though we will apply the prescription \eqref{BasicTr} to twistor representatives and not to cohomology classes, this is sufficient for the purposes of this work, which are simply to use twistor theory as a tool to show how to relate different solutions via complex coordinate transformations.

\section{Complex transformations}

For most of our examples of interest, we will need the following identity:
\begin{proposition}
Let $\alpha_{A'},\beta_{A'}$ be two arbitrary 
(non-proportional) spinor fields, and let $r,s$ be positive integers. 
Then:
\begin{align}
\frac{1}{2\pi\i}\oint_{\Gamma}\frac{\pi_{A'_1}...\pi_{A'_{2h}}\pi_{B'}\d\pi^{B'}}{(\alpha^{A'}\pi_{A'})^{r}(\beta^{A'}\pi_{A'})^{s}} 
= \frac{k}{(\alpha_{A'}\beta^{A'})^{2h+1}}
 \beta_{(A'_{1}}...\beta_{A'_{r-1}}\alpha_{A'_{r}}...\alpha_{A'_{2h})}
 \label{identity}
\end{align}
where $k$ is a constant, $2h=r+s-2$, and
the contour $\Gamma$ separates the poles at $\pi_{A'}=\alpha_{A'}$ and $\pi_{A'}=\beta_{A'}$.
\end{proposition}
A simple way to show \eqref{identity} is to contract the left hand side 
with $n$ factors of $\alpha^{A'}$ and $2h-n$ factors of $\beta^{A'}$, 
and then deduce that the integral will be 
equal to $(\alpha_{A'}\beta^{A'})^{-1}$ if $n=r-1$ and zero otherwise; 
the right hand side then follows straightforwardly.

\subsection{Constant, elementary, and momentum states}

An elementary state in twistor theory \cite{PMC73, PR2} is a zero-rest-mass field generated by a twistor function of the form
\begin{align}
 f(Z) = \frac{(C_{\alpha}Z^{\alpha})^{l}(D_{\alpha}Z^{\alpha})^{m}}
 {(A_{\alpha}Z^{\alpha})^{r}(B_{\alpha}Z^{\alpha})^{s}},
 \label{ES}
\end{align}
for some $A_{\alpha},...,D_{\alpha}$, where $l,m,r,s$ are non-negative integers. 
The relevance of these functions comes from their utility as an 
alternative basis to momentum eigenstates 
\cite[Section 4]{PMC73}, \cite{HuggettTod}, \cite{Hodges}.
We will focus on the case $l=m=0$, 
so that \eqref{ES} takes the form 
\begin{align}
f(Z)=[\chi(Z)]^{-1}, \qquad 
\chi(Z)=(A_{\alpha}Z^{\alpha})^{r}(B_{\alpha}Z^{\alpha})^{s}.
\label{ES2}
\end{align}

Consider the simple case
\begin{align}
 \chi(Z)=(Z^{2})^{r}(Z^{3})^{s}. \label{Constantfields}
\end{align}
The singular region of $f=\chi^{-1}$ is the 
algebraic set $\{\chi=0\}$, which consists of two 
parallel planes $\mathbb{A}=\{Z^2=0\}$, $\mathbb{B}=\{Z^{3}=0\}$.
These are in fact two fibers of the fibration $\PT\to\CP^1$ 
(as such, they do not intersect). 
The Penrose transform of $f=\chi^{-1}$ is a particular case 
of \eqref{identity}, with $\alpha^{A'}=o^{A'}$, $\beta^{A'}=\iota^{A'}$ 
(see appendix \ref{Appendix} for our conventions).
Thus we immediately obtain the 
constant spinor field
$\varphi_{A'_1...A'_{2h}}=k \iota_{(A'_{1}}...\iota_{A'_{r-1}}o_{A'_{r}}
...o_{A'_{2h})}$, with $2h=r+s-2$. 
For example, for $r=s=2$, the corresponding (self-dual) Maxwell field is (see eq. \eqref{spin1} below)
\begin{align}
\mathcal{F}=\d{t}\wedge\d{z}+\i\d{x}\wedge\d{y}, 
\label{constantMaxwell}
\end{align}
which (assuming $t,x,y,z$ to be real) is a constant electric field 
in the $z$ direction.

Now, writing \eqref{Constantfields} as in \eqref{ES2} 
with $A_{\alpha}=(0,0,1,0)$, $B_{\alpha}=(0,0,0,1)$,
we apply an ${\rm SL}(4,\mathbb{C})$ transformation 
$Z^{\alpha}\mapsto Z'^{\alpha}=T^{\alpha}{}_{\beta}Z^{\beta}$
and put  
\begin{align}
 \chi'(Z) := \chi(Z') = (A'_{\alpha}Z^{\alpha})^{r}(B'_{\alpha}Z^{\alpha})^{s},
 \label{tfhopfions}
\end{align}
where $A'_{\beta}=A_{\alpha}T^{\alpha}{}_{\beta}=
(\nu_{0'B},\tilde\theta_{0'}{}^{B'})$, 
$B'_{\beta}=B_{\alpha}T^{\alpha}{}_{\beta}=(\nu_{1'B},\tilde\theta_{1'}{}^{B'})$ 
($T^{\alpha}{}_{\beta}$ is given explicitly by \eqref{genericT}).
We see that $A'_{\alpha}, B'_{\alpha}$ are only sensitive to the parts of $T^{\alpha}{}_{\beta}$ corresponding to right Lorentz rotations and dilations (contained in $\tilde\theta_{A'}{}^{B'}$) and special conformal transformations (contained in $\nu_{A'B}$). 
The singular set of $f'=\chi'^{-1}$ is again given by two planes 
in $\PT$, $\mathbb{A'}=\{A'_{\alpha}Z^{\alpha}=0\}$, 
$\mathbb{B'}=\{B'_{\alpha}Z^{\alpha}=0\}$,
but now the planes intersect. This intersection is a twistor line, 
$\mathbb{A'}\cap\mathbb{B'}=L_{q}$, where, putting 
$A'_{\alpha}=(a_{A},\tilde{a}^{A'})$, 
$B'_{\alpha}=(b_{A},\tilde{b}^{A'})$, 
the point $q\in\CM$ is given by 
\begin{align}
q^{AA'}=
\frac{\i}{(a_Bb^B)}(b^A\tilde{a}^{A'}-a^A\tilde{b}^{A'}) = 
\frac{2\i}{(\nu_{c}\nu^{c})} \nu^{AB'}\tilde\theta_{B'}{}^{A'}
\label{pointq}
\end{align}
(we assume $\nu^{a}$ to be non-null).
The Penrose transform of $f'=\chi'^{-1}$ is again a particular 
case of \eqref{identity}, where now 
$\alpha^{A'}=\i x^{AA'}a_{A}+\tilde{a}^{A'}$, 
$\beta^{A'}=\i x^{AA'}b_{A}+\tilde{b}^{A'}$.
The zero-rest-mass field is then given by the right hand side of \eqref{identity}, with
\begin{align}
\alpha_{A'}\beta^{A'} = k' (x_a-q_a)(x^a-q^a)
 \label{hopfions}
\end{align}
for some constant $k'$.
The fields represented by the right hand side of \eqref{identity} with 
\eqref{hopfions} are called `spin-$h$ hopfions' 
or `knotted fields' \cite{Bouwmeester, Bouwmeester2}. 
The name comes from the fact that the principal spinors 
$\alpha_{A'},\beta_{A'}$ define in this case Robinson congruences, 
which are in turn related to the Hopf fibration 
(cf. \cite[section 6.2]{PR2}).

From the observation made around eq. \eqref{BasicTr},
we deduce that spin-$h$ hopfions/knotted fields 
can be obtained via complex 
conformal transformations of constant fields. 
We notice that, for the case of null fields in electromagnetism, 
a complex transformation from a constant, null Maxwell 
field to a null electromagnetic hopfion was already given in 
\cite{BB}, \cite{Hoyos}, the null condition being essential. 
Our approach in this work shows that the complex 
transformation is valid for fields of arbitrary spin and arbitrary 
algebraic type, and it is just a particular example of the general 
framework given around eq. \eqref{BasicTr}.

\subsubsection*{An intuitive interpretation}

Even though the above result that parallel spinors, which 
are simply constant fields on space-time, and spin-$h$ 
hopfions/knotted fields, which have a quite complicated topological 
structure, are related by 
a complex conformal transformation may be 
difficult to anticipate if one only looks at their space-time description, 
from the twistor perspective we can get a fairly simple intuitive 
understanding of this phenomenon. 
Constant fields can be generated by two parallel planes in $\PT$ 
($\{Z^2=0\}$ and $\{Z^3=0\}$), 
while spin-$h$ hopfions can be generated by two intersecting 
planes\footnote{Indeed, this elementary observation was one of 
our basic motivations for this work.}. 
In the former case, the planes do not intersect because we removed 
the line $\textbf{I}$ from $\CP^3$ (which corresponds to infinity in 
space-time), 
while in the latter, the planes intersect in the twistor line $L_q$ 
corresponding to a point $q$ in space-time. 
If we think of the two parallel planes as ``intersecting at infinity'', 
then the above two twistor configurations are equivalent, 
so the corresponding physical configurations must be appropriately 
equivalent as well. 
It is also clear that the transformation must really be conformal, 
since Poincar\'e transformations preserve the line ${\bf I}$ 
(i.e. the infinity twistor $I_{\alpha\beta}$).

A little more formally, recall that in order for conformal 
transformations to be well-defined everywhere in space-time, 
we must consider 
compactified Minkowski space $\CM^{\sharp}$. 
As mentioned in section \ref{Sec:Preliminaries},
the twistor space of $\CM^{\sharp}$ (which is $\CP^3$)
does contain ${\bf I}$, and the planes $\{Z^2=0\}$ and $\{Z^3=0\}$ 
intersect precisely in ${\bf I}$.
The special conformal transformation relating parallel spinors and 
spin-$h$ hopfions interchanges the line ${\bf I}$ (defined by 
$Z^2=Z^3=0$) and the line $L_{q}$ (defined by $Z'^2=Z'^3=0$),
or equivalently, it interchanges (via conformal inversion)
the point $q\in\CM^{\sharp}$ with the vertex $I$ of the light-cone at 
infinity (which is also a point in $\CM^{\sharp}$).
In other words, if we interpret the point $q$ as the location of the 
``source'' of the spin-$h$ hopfion, we see that 
the source of a constant field is a point at infinity.

\subsubsection*{Momentum eigenstates}

Consider now twistor functions of the form
\begin{align} 
 f(Z) = \frac{\exp(C_{\alpha}Z^{\alpha}/B_{\alpha}Z^{\alpha})}
 {(A_{\alpha}Z^{\alpha})(B_{\alpha}Z^{\alpha})^{2h+1}}
 \label{MomentumStates}
\end{align}
where $A_{\alpha}=(0,\tilde{a}^{A'})$, $B_{\alpha}=(0,\tilde{b}^{A'})$, 
$C_{\alpha}=(c_{A},0)$. 
The corresponding zero-rest-mass fields are momentum eigenstates 
(or plane waves; see \cite[section 4.4]{PMC73}): 
choosing $\tilde{a}_{A'}\tilde{b}^{A'}=1$, 
defining $k_{a}:=c_{A}\tilde{a}_{A'}$, and applying a slight variation 
of formula \eqref{identity}, we get the null fields 
\begin{align}
 \varphi_{A'_1...A'_{2h}} = e^{\i x^{a}k_{a}}\tilde{a}_{A'_1}...
 \tilde{a}_{A'_{2h}}.
 \label{waves}
\end{align}
Now perform an arbitrary linear transformation $Z^{\alpha}\mapsto 
Z'^{\alpha}=T^{\alpha}{}_{\beta}Z^{\beta}$, and define 
$A'_{\beta}=A_{\alpha}T^{\alpha}{}_{\beta}=(a'_A,\tilde{a}'^{A'})$, 
and similarly for $B'_{\beta}, C'_{\beta}$.
Let $f'(Z):=f(Z')$. Then the Penrose transform of $f'$ gives
\begin{align}
 \varphi'_{A'_1...A'_{2h}} =
 \frac{\exp\left[\frac{\alpha_{A'}\gamma^{A'}}{\alpha_{B'}\beta^{B'}}\right]}
 {(\alpha_{C'}\beta^{C'})^{2h+1}}\alpha_{A'_1}...\alpha_{A'_{2h}}.
 \label{hopfionswaves}
\end{align}
where $\alpha^{A'}=\i x^{AA'}a'_{A}+\tilde{a}'^{A'}$, etc.
The prefactor can be written as 
$$\tfrac{1}{(\alpha_{C'}\beta^{C'})^{2h+1}}\exp\left[\tfrac{\alpha_{A'}\gamma^{A'}}{\alpha_{B'}\beta^{B'}}\right]=
\tfrac{c_1}{|x-q|^{2(2h+1)}}\exp\left[c_2\tfrac{|x-p|^2}{|x-q|^2}\right]$$
for some constants $c_1,c_2$, and some fixed points $q^a,p^a$ 
defined analogously to \eqref{pointq}. 
(We also put $|x-q|^{2}\equiv (x_{a}-q_{a})(x^{a}-q^{a})$, etc.)
Thus, a complex conformal transformation of a plane wave 
\eqref{waves} produces a new null, zero-rest-mass field 
\eqref{hopfionswaves} with hopfion/knotted-like features.
We notice that for the electromagnetic case ($h=1$), 
similar results were obtained in \cite{Hoyos} (see also \cite{BB}).
The generalization \eqref{hopfionswaves} to arbitrary spin 
appears to be new. 
The $h=2$ case might exhibit some interesting physical features, 
as it transforms a plane gravitational wave to a hopfion/knotted-like 
gravitational wave (which is different from the gravitational hopfion 
considered in \cite{Bouwmeester, Bouwmeester2}).

\subsection{Linearized black holes}

By `linearized black hole' we mean a spin 2 field ($h=2$ in \eqref{ZRMF}) in $\CM$ which formally looks the same as the Weyl curvature spinor of a black hole (Petrov type D) solution 
\cite{PMC73}. 
As explained in \cite{PMC73}, such fields are generated by twistor 
functions of the form 
\begin{align}
 f(Z) = [\chi(Z)]^{-(h+1)}, \qquad
 \chi(Z)=Q_{\alpha\beta}Z^{\alpha}Z^{\beta},
 \label{LinearizedBH}
\end{align}
where $h=2$, and $Q_{\alpha\beta}$ cannot be written as 
a product $A_{(\alpha}B_{\beta)}$ of only two twistors. 
The $h=1$ case of \eqref{LinearizedBH} describes electromagnetic 
analogues such as the Coulomb field, the ``magic'' \cite{LyndenBell}
(or ``root-Kerr'' \cite{ArkaniHamed}) field, or others (see examples below). 
We will however leave $h$ in \eqref{LinearizedBH} arbitrary, 
so our construction also applies to higher/lower spin analogues.

The facts that $Q_{\alpha\beta}\neq A_{(\alpha}B_{\beta)}$ and that any conformal transformation of $A_{(\alpha}B_{\beta)}Z^{\alpha}Z^{\beta}$ gives $A'_{(\alpha}B'_{\beta)}Z^{\alpha}Z^{\beta}$ suggest that linearized black holes (and higher/lower spin analogues) cannot be obtained from complex conformal transformations of elementary states. 
However, we should be more careful since these are facts about representatives and not about cohomology classes. In other words, one would need to prove that the cohomology classes of \eqref{LinearizedBH} and \eqref{ES2} cannot be connected by a linear transformation of the twistor variables.
We will not attempt to do this, and will simply study the case \eqref{LinearizedBH} separately.

For any (non-negative) integer $h$, the zero-rest-mass fields 
generated by \eqref{LinearizedBH} can be obtained as a particular 
case of formula \eqref{identity}, with $r=s=h+1$. 
This is because on a generic twistor line $L_x$, using the incidence relation \eqref{IR} we get
$\chi|_{L_x}=K^{A'B'}\pi_{A'}\pi_{B'}$ for some symmetric $K^{A'B'}$, 
which can always be decomposed into principal spinors as
$K^{A'B'}=\alpha^{(A'}\beta^{B')}$, and, 
assuming the generic case $K^{A'B'}K_{A'B'}\neq0$, 
we have $\alpha_{A'}\beta^{A'}\neq0$. 
The spinors $\alpha_{A'}, \beta_{A'}$ contain the information of the 
roots of the second order homogeneous polynomial 
$\chi|_{L_x}=K^{A'B'}\pi_{A'}\pi_{B'}$. 
More explicitly, in terms of a coordinate 
$\zeta=\frac{\pi_{1'}}{\pi_{0'}}$ on the Riemann sphere of $x$, 
we have $\chi|_{L_x}=(\pi_{0'})^2(A\zeta^2+2B\zeta+C)$, 
where $A=K^{1'1'}$, $B=K^{0'1'}$, $C=K^{0'0'}$.
The roots are then
\begin{align}
 \zeta_{\pm} = \tfrac{1}{A}(-B\pm\Delta), 
 \qquad \Delta:=\sqrt{B^2-AC}.
 \label{roots}
\end{align}
Putting $\alpha^{A'}=\sqrt{A}(o^{A'}+\zeta_{+}\iota^{A'})$, 
$\beta^{A'}=\sqrt{A}(o^{A'}+\zeta_{-}\iota^{A'})$, we get 
$K^{A'B'}=\alpha^{(A'}\beta^{B')}$ as required.
We also see that $\alpha_{A'}\beta^{A'}=-2\Delta$.
Finally, defining a spin frame 
$\hat{\alpha}_{A'}=\frac{1}{(\alpha_{B'}\beta^{B'})^{1/2}}\alpha_{A'}$,
$\hat{\beta}_{A'}=\frac{1}{(\alpha_{B'}\beta^{B'})^{1/2}}\beta_{A'}$, 
we can express the field generated by \eqref{LinearizedBH} as 
\begin{align}
\varphi_{A'_1...A'_{2h}} = \frac{k}{\Delta^{h+1}}
\hat\alpha_{(A'_1}...\hat\alpha_{A'_h}\hat\beta_{A'_{h+1}}...\hat\beta_{A'_{2h})}
\label{Lbhspinh2}
\end{align}
where $\hat\alpha_{A'}\hat\beta^{A'}=1$ and we redefined the 
constant $k$. From the invariant expression
\begin{align}
 \varphi^{A'_1...A'_{2h}}\varphi_{A'_1...A'_{2h}} \propto 
 \frac{1}{\Delta^{2(h+1)}},
\end{align}
we see that the field is not null, and that it is singular at $\Delta=0$. 
This will be useful for physical interpretation.

The case $h=1$ (Maxwell fields) has a simple description in tensor 
terms: defining $\mathcal{F}_{ab}=\varphi_{A'B'}\epsilon_{AB}$,
a calculation gives
\begin{align}
 \mathcal{F} = \frac{k}{\Delta^{3}}\left[ 
 \tfrac{(A-C)}{2}(\d{t}\wedge\d{x}+\i\d{y}\wedge\d{z}) 
 + B(\d{t}\wedge\d{z}+\i\d{x}\wedge\d{y})
 +\tfrac{(A+C)}{2}(\d{z}\wedge\d{x}+\i\d{y}\wedge\d{t})
 \right].
 \label{spin1}
\end{align}

\subsubsection*{Schwarzschild and Pleba\'nski-Demia\'nski}

Consider the twistor function \eqref{LinearizedBH}  with
\begin{align}
 \chi(Z) = Z^{0}Z^{3} - Z^{1}Z^{2}.
 \label{BHs}
\end{align}
On a generic twistor line $L_x$, we get $A=\frac{x+\i y}{\sqrt{2}}$, 
$B=\frac{z}{\sqrt{2}}$, $C=-\frac{(x-\i y)}{\sqrt{2}}$ (we omit an overall factor of $\i$).
The function $\Delta$ in \eqref{roots} is
\begin{align}
\Delta= \tfrac{1}{\sqrt{2}}r, \qquad r:=\sqrt{x^2+y^2+z^2}
\end{align}
and the roots are $\zeta_{\pm}=\frac{-z\pm r}{x+\i y}$.
The singularity region $\Delta=0$ of the field \eqref{Lbhspinh2} is then $x=y=z=0$ ($t$ arbitrary), which we can interpret as 
Coulomb/Schwarzschild-like behavior.
For example, for $h=1$ and $h=2$:
\begin{align}
\mathcal{F} = \frac{k}{r^2}\left[\d{t}\wedge\d{r} 
- \i r^2\sin\theta\d\phi\wedge\d\theta\right],
\qquad
\varphi_{A'B'C'D'} = \frac{k}{r^{3}}\hat\alpha_{(A'}\hat\alpha_{B'}
\hat\beta_{C'}\hat\beta_{D')}
\end{align}
where $(r,\theta,\phi)$ are standard spherical coordinates, 
defined by $x+\i y=r\sin\theta e^{\i\phi}$, $z=r\cos\theta$. 
The field $\mathcal{F}$ is precisely the (self-dual) Coulomb field, 
while the spin 2 field is the linearized Schwarzschild solution.

Now consider a linear transformation 
$Z^{\alpha}\mapsto Z'^{\alpha}=T^{\alpha}{}_{\beta}Z^{\beta}$, with
\begin{align}
T^{\alpha}{}_{\beta} = 
\left(\begin{matrix} 
1 & 0 & \frac{c}{\sqrt{2}} & 0 \\
 0 & 1 & 0 & -\frac{c}{\sqrt{2}} \\
 -\frac{1}{c\sqrt{2}} & 0 & \frac{1}{2} & 0 \\
 0 & -\frac{1}{c\sqrt{2}} & 0 & \frac{1}{2}
\end{matrix}\right)
\label{TSPD}
\end{align}
for some $c\in\mathbb{C}$, $c\neq0$. One can check that 
$\det(T^{\alpha}{}_{\beta})=1$, 
so $T^{\alpha}{}_{\beta}\in{\rm SL}(4,\mathbb{C})$.
Following the prescription \eqref{BasicTr} with \eqref{LinearizedBH} 
and \eqref{BHs}, we get
\begin{align}
\chi'(Z):=\chi(Z') = \tfrac{\sqrt{2}}{c} 
\left( Z^{0}Z^{1}+\tfrac{c^2}{2}Z^{2}Z^{3} \right).
\label{chiPD}
\end{align}
After some straightforward calculations, the new $\Delta$ \eqref{roots}, now denoted $\Delta'$, is
\begin{align}
\Delta'=\tfrac{1}{4}\sqrt{(x_ax^a-c^2)^2-4c^2(x^2+y^2)}.
 \label{DeltaPD}
\end{align}
In order to have an interpretation of the new field, we must 
analyze the set of points on space-time where $\Delta'=0$.
To this end, we separate $c$ into real and imaginary parts as 
\begin{align}
c=a+\i b, \label{PDparameter}
\end{align}
with $a,b$ real. Writing also $(4\Delta')^2=R+\i I$ (with $R,I$ real), 
we get
\begin{equation}
\begin{aligned}
 R = {}& (x_ax^a)^{2}+a^4+b^4-6a^2b^2
 +2(a^2-b^2)(z^2-t^2-x^2-y^2), \\
 I ={}& 4ab(a^2-b^2+z^2-t^2-x^2-y^2), 
\end{aligned}
\label{Im}
\end{equation}
so $\Delta'=0$ iff $R=I=0$. 
The condition $I=0$ gives $t^2-z^2+x^2+y^2=a^2-b^2$. Replacing 
in $R=0$, we get also $t^2-z^2-(x^2+y^2)=\pm(a^2+b^2)$.
The $+$ sign leads to $t^2-z^2=a^2$ and $x^2+y^2=-b^2$, while the 
$-$ sign leads to $z^2-t^2=b^2$ and $x^2+y^2=a^2$.
Assuming the generic case $a\neq0$, $b\neq0$ (see the next 
example for other cases), we find:
\begin{align}
 \Delta'=0 \qquad \Leftrightarrow \qquad 
 z^2-t^2=b^2 \quad \text{and} \quad x^2+y^2 = a^2.
\end{align}
This is exactly the singular structure of the Pleba\'nski-Demia\'nski field \cite{PD}: two accelerating ring singularities $x^2+y^2=a^2$, each moving on one branch of the hyperbola $z^2-t^2=b^2$. 

In order to interpret \eqref{TSPD} in space-time terms, we 
express it as a composition of the basic transformations 
\eqref{genericT}-\eqref{Tx}. We find:
\begin{align}
 T^{\alpha}{}_{\beta} = 
 S^{\alpha}{}_{\gamma}U^{\gamma}{}_{\beta}, 
 \qquad
S^{\alpha}{}_{\gamma} = \left( \begin{matrix}
\delta^{A}_{C} & 0 \\ \frac{1}{c^2} \tau_{A'C} & \delta_{A'}^{C'}
\end{matrix}\right), \quad
 U^{\gamma}{}_{\beta} = \left( \begin{matrix}
\delta^{C}_{B} & \tau^{CB'} \\ 0 & \delta_{C'}^{B'}
\end{matrix}\right),
\end{align}
where the components of $\tau^{AB'}$ are 
$\tau^{00'}=\frac{c}{\sqrt{2}}=-\tau^{11'}$, 
$\tau^{01'}=\tau^{10'}=0$ (see appendix \ref{Appendix} 
for some useful identities).
Using \eqref{Tx}, we see that $U^{\alpha}{}_{\beta}$ corresponds 
to a translation along the vector field $\xi^{a}$
while $S^{\alpha}{}_{\beta}$ is a special conformal transformation 
along $s^{a}=\frac{1}{c^2}\xi^{a}$, where 
(recalling the definition \eqref{PDparameter})
\begin{align}
 \xi^{a}=(0,0,0,-\i a + b). \label{PDVF}
\end{align}
Summarizing:
\begin{proposition}\label{Prop:SPD}
The Pleba\'nski-Demia\'nski field 
can be obtained from the Schwarzschild field (for any spin $h$, in 
particular for the linearized black hole solutions) 
by a complex translation along $\xi^{a}$ followed by a
complex special conformal transformation along 
$s^{a}=\frac{1}{(a+\i b)^2}\xi^{a}$, where 
$\xi^{a}$ is given by \eqref{PDVF}.
\end{proposition}

This provides an interpretation for the transformation for 
Maxwell fields mentioned by Pleba\'nski and Demia\'nski in 
\cite[Eq. (4.65)]{PD}. 
To have some intuition about the appearance of two objects ``out of one'', see the Conclusions \ref{Sec:Conclusions}.

\subsubsection*{Kerr and the C-metric}

Let $a$ be a real parameter, and consider
\begin{align}
 \chi(Z) = Z^{0}Z^{3} - Z^{1}Z^{2} + \sqrt{2} a \: Z^2Z^3. 
 \label{chiKerr}
\end{align}
Note that this function can be obtained from \eqref{BHs} by a 
linear transformation \eqref{LinearT}-\eqref{genericT}
corresponding to a translation: this is the twistor version of the 
(linearized) Newman-Janis shift. We will, however, analyze this 
case independently of \eqref{BHs}.
On twistor lines, we find $A=\frac{x+\i y}{\sqrt{2}}$, 
$B=\frac{z-\i a}{\sqrt{2}}$, $C=-\frac{(x-\i y)}{\sqrt{2}}$, which gives
\begin{align}
\Delta=\tfrac{1}{\sqrt{2}}\rc, \qquad 
\rc:=\sqrt{x^2+y^2+(z-\i a)^2} = r-\i a z/r,
\end{align}
where $r$ is defined to be the real part of $\rc$.
The singularity region $\Delta=0$ of the fields \eqref{Lbhspinh2} is now $x^2+y^2=a^2$, $z=0$ ($t$ arbitrary). 
This ring singularity allows us to associate \eqref{Lbhspinh2} 
in this case to the (linearized) Kerr field and its higher/lower 
spin analogues. For example, introducing a spheroidal coordinate 
system 
$(r,\theta,\phi)$ by $x+\i y=\sqrt{r^2+a^2}\sin\theta e^{\i\phi}$, 
$z=r\cos\theta$, we find for the spin 1 and 2 cases:
\begin{align}
\mathcal{F} ={}& \frac{k}{(r-\i a\cos\theta)^{2}}
\left[ \d{t}\wedge(\d{r}+\i a\sin\theta\d\theta)-\sin\theta\d\phi\wedge
(a\sin\theta\d{r}+\i(r^2+a^2)\d\theta)\right], \label{rootKerr} \\
\varphi_{A'B'C'D'} ={}& \frac{k}{(r-\i a\cos\theta)^{3}}\hat\alpha_{(A'}\hat\alpha_{B'}\hat\beta_{C'}\hat\beta_{D')}. \label{linKerr}
\end{align}
The field \eqref{rootKerr} is the root-Kerr (or magic) solution \cite{ArkaniHamed}, while \eqref{linKerr} is the linearized Kerr solution. So we can interpret $a$ as an angular momentum parameter.

Consider now the transformation
$Z^{\alpha}\mapsto Z'^{\alpha}=T^{\alpha}{}_{\beta}Z^{\beta}$ with
\begin{align}
T^{\alpha}{}_{\beta} = 
\left(\begin{matrix} 
 \lambda & 0 & 0 & 0 \\
 0 & \lambda & 0 & 0 \\
 -\frac{\lambda}{a\sqrt{2}} & 0 & \frac{1}{\lambda} & 0 \\
 0 & \frac{\lambda}{a\sqrt{2}} & 0 & \frac{1}{\lambda}
\end{matrix}\right)
\label{TKC}
\end{align}
where $\lambda\neq 0$ is a complex parameter. 
Following the prescription \eqref{BasicTr} with \eqref{LinearizedBH} 
and \eqref{chiKerr}, we find
\begin{align}
\chi'(Z):=\chi(Z') = \tfrac{\sqrt{2}}{c}
\left(Z^{0}Z^{1}+\tfrac{c^2}{2}Z^{2}Z^{3}\right), \qquad 
c=\frac{2a}{\lambda^2}.
\label{chiC}
\end{align}
So the new function $\chi'$ is formally the same as \eqref{chiPD},
and the new $\Delta'$ is again given by \eqref{DeltaPD} 
(although the parameter $c$ is now related to a different conformal 
transformation \eqref{TKC}-\eqref{chiC}).
To interpret the new field, we analyze the set of points where 
$\Delta'=0$. This was already done below eq. \eqref{Im} 
when $c$ is genuinely complex, in which case the new solution 
is the Pleba\'nski-Demia\'nski field. 
So it remains to analyze the cases $c$ real or purely imaginary:
\begin{itemize} 
\item Suppose first $c$ purely imaginary, $c\equiv \i \alpha^{-1}$, 
$\alpha\in\mathbb{R}$. Then 
$(4\Delta')^2=(x_{a}x^{a}+\alpha^{-2})^{2}+4\alpha^{-2}(x^2+y^2)$, 
so $\Delta'=0$ iff $x^2+y^2=0=x_{a}x^{a}+\alpha^{-2}$. 
This gives $x=y=0$, $z^2-t^2=\alpha^{-2}$. These are two points moving each on one branch of the hyperbola $z^2-t^2=\alpha^{-2}$: this is the {\em C-metric field}, with acceleration parameter $\alpha$.
\item Suppose now $c$ real (for concreteness assume $c>0$). Then we can write the equation $\Delta'=0$ as $(x^2+y^2+z^2+c^2-t^2)^2=4c^2(x^2+y^2)$. For $t=0$, we have a ring singularity $x^2+y^2=c^2$, $z=0$. For fixed $t\neq0$, the equation describes a torus, where $t$ is the radius of the tube and $c$ is the distance from the center of the torus to the center of the tube. As time progresses, the torus evolves through the three possible tori\footnote{See e.g. \href{https://en.wikipedia.org/wiki/Torus}{\texttt{https://en.wikipedia.org/wiki/Torus}}.}: the standard ``ring torus'' for $t<c$, a ``horn torus'' (no hole) at $t=c$, and a ``spindle torus'' (self-intersecting) for $t>c$. This toroidal singularity is quite peculiar, and we are not aware of a non-linear solution in general relativity that can be associated to this field. 
We note however that this singularity has also been described in \cite{Kassandrov}.
\end{itemize}

Finally, we need to interpret \eqref{TKC} in space-time terms. 
To this end, we note that 
\begin{align}
 T^{\alpha}{}_{\beta} = 
 D^{\alpha}{}_{\gamma}S^{\gamma}{}_{\beta}, 
 \qquad
D^{\alpha}{}_{\gamma} = \left( \begin{matrix}
\lambda\delta^{A}_{C} & 0 \\ 0 & \lambda^{-1}\delta_{A'}^{C'}
\end{matrix}\right), \quad
 S^{\gamma}{}_{\beta} = \left( \begin{matrix}
\delta^{C}_{B} & 0 \\ \nu_{C'B} & \delta_{C'}^{B'}
\end{matrix}\right),
\end{align}
where the components of $\nu_{A'B}$ are 
$\nu_{0'0}=-\frac{\lambda^2}{a\sqrt{2}}=-\nu_{1'1}$, 
$\nu_{0'1}=\nu_{1'0}=0$. 
Using \eqref{Tx}, we see that $D^{\alpha}{}_{\beta}$ is a 
dilation with parameter $\lambda$, and $S^{\alpha}{}_{\beta}$ is a 
special conformal transformation along the vector field 
\begin{align}
 s^{a} = (0,0,0,\tfrac{\lambda^2}{2\i a}). \label{VFKC}
\end{align}
To summarize:
\begin{proposition}
The conformal transformation \eqref{TKC}, which consists of a 
complex special conformal transformation along \eqref{VFKC} 
followed by a complex dilation with parameter $\lambda$, 
maps the Kerr field (for any spin $h$, in particular for the linearized 
black hole solutions) to: 
$(i)$ the Pleba\'nski-Demia\'nski field if $\lambda^2$ is genuinely complex,
$(ii)$ the C-metric field if $\lambda^2$ is purely imaginary, 
$(iii)$ a toroidal singularity if $\lambda^2$ is real.
\end{proposition}

Note that the Schwarzschild field can also be transformed to the C-metric field and to the toroidal singularity (by assuming $a=0$ or $b=0$ in \eqref{PDparameter}). On the other hand,
given that the Pleba\'nski-Demia\'nski field can be obtained 
from the Schwarzschild field by a translation followed by a 
special conformal transformation, and that the Kerr field is itself 
obtained from a translation of Schwarzschild, one might think that 
to go from Kerr to Pleba\'nski-Demia\'nski one only needs 
a special conformal transformation. 
But the above result shows that a complex dilation is also needed.
In view of the definition of $c$ in \eqref{chiC}, we see that the 
action of the dilation is to effectively complexify the angular 
momentum parameter (so that it becomes $c$).

\subsection{Spherical scalar waves}

As a final example of complex conformal transformations, 
we can try to combine the twistor functions of 
linearized black holes \eqref{LinearizedBH} with the 
ones of plane waves \eqref{MomentumStates}. That is, consider 
\begin{align}
 f(Z) = \frac{\exp(C_{\alpha}Z^{\alpha}/B_{\alpha}Z^{\alpha})}
 {(Q_{\alpha\beta}Z^{\alpha}Z^{\beta})^{h+1}}
 \label{SphericalWaves}
\end{align}
where $B_{\alpha}=(0,b^{A'})$, $C_{\alpha}=(c_{A},0)$, 
and $Q_{\alpha\beta}$ cannot be expressed as a product of 
only two twistors. 
For spin other than zero (i.e. $h>0$), the calculation of the contour 
integrals involved in the zero-rest-mass fields associated to 
\eqref{SphericalWaves} is quite involved. 
We will, for simplicity, restrict ourselves to the scalar case $h=0$.
A calculation shows that 
the Penrose transform of the twistor function 
\eqref{SphericalWaves} with $h=0$ is 
\begin{align}
\varphi(x^{a}) = \frac{1}{\alpha_{A'}\beta^{A'}}
\exp\left[ \i \frac{x^{AA'}c_{A}\alpha_{A'}}{\alpha_{B'}b^{B'}}\right]
\end{align}
(in this and the following expressions we will omit irrelevant overall numerical constants),
where $\alpha_{A'},\beta_{A'}$ are the spinor fields defined by 
$(Q_{\alpha\beta}Z^{\alpha}Z^{\beta})|_{L_x}=
\alpha^{A'}\beta^{B'}\pi_{A'}\pi_{B'}$.
Choosing $c_{A}=o_{A}$, 
$b^{A'}=\frac{1}{\sqrt{2}}\iota^{A'}$, and recalling the definitions 
\eqref{roots} of $\Delta$ and $\zeta_{\pm}$, we get 
\begin{align}
\varphi(x^{a}) = \frac{e^{\i(t+z+\zeta_{+}(x+\i y))}}{\Delta}.
\end{align}
For example, choosing $Q_{\alpha\beta}$, $Q'_{\alpha\beta}$ 
to be given by \eqref{BHs} and \eqref{chiPD} respectively, 
we get the scalar waves
\begin{subequations}
\begin{align}
\varphi(x^{a}) ={}& \frac{1}{r}\exp\left( \i(t+r)\right)
\label{SW2}, \\
\varphi'(x^{a}) ={}& \frac{1}{\Delta'}
\exp\left[\i(t+z)+\tfrac{\i}{2(t-z)}
\left(-t^2+z^2-x^2-y^2+c^2+ \Delta'\right) \right] \label{SW3}
\end{align}
\end{subequations}
where $r=\sqrt{x^2+y^2+z^2}$ and $\Delta'$ is defined in 
\eqref{DeltaPD}. We see that \eqref{SW2} represents 
a spherical scalar wave, and \eqref{SW3} is a much more complicated configuration, but the two solutions \eqref{SW2}-\eqref{SW3} are related by the complex conformal transformation mentioned in proposition \ref{Prop:SPD}.

\section{Comments on non-linear fields}

As emphasized by Newman \cite{Newman88}, Flaherty \cite{Flaherty}, and others, the fact that complex coordinate 
transformations produce new solutions to real field equations can 
be understood (assuming real-analyticity) via consideration 
of holomorphic extensions, holomorphic coordinate transformations, and the imposition of new reality conditions. The coordinate transformations should also preserve some additional structure, e.g. the Minkowski metric, or the conformal structure in the current manuscript \footnote{Otherwise, all (say, non-null) Maxwell fields 
could be deemed as ``equivalent'', since, given any two of them, one can be mapped to the other by the transformation that takes the Darboux coordinates of the first to the Darboux coordinates of the other. Mathematically, 
this is the statement that all symplectic forms are locally 
equivalent.}.

While the above works well for linear theories in (conformally) flat space-time and it has been the topic of this manuscript, the understanding of the fully non-linear Newman-Janis shift \cite{NewmanJanis} in general relativity is much less satisfactory. 
In general relativity, one would wish to preserve the Einstein equations, i.e., to map a solution to another solution.
This is guaranteed to be the case if one considers 
holomorphic extensions. 
The non-linear Newman-Janis shift, however, does not 
correspond to a holomorphic extension, and the fact that it produces a new solution does not seem to be an automatic consequence of the procedure.
What is more, the transformation between Schwarzschild and Kerr 
{\em cannot} be holomorphic (at least in the above sense of analytic continuation), since, as explained by Newman \cite{Newman88}, the solutions have different numbers of holomorphic Killing vectors.

As is well-known, the main ambiguity in the non-linear Newman-Janis shift is in the way in which some functions in the metric must be complexified. In particular, the function $\frac{2m}{r}$ must be replaced by $\frac{m}{r}+\frac{m}{\bar{r}}$ in order for the trick to work. More generally, see \cite{AdamoNewman}, the idea is that a function $f(r)$ must be replaced by a function $F(r,\bar{r})$ that reduces to $f(r)$ on the real slice. We can actually use this to give a very simple version of the trick, as follows. 
(We are not aware that this form of the trick has been given before.) 
Consider a complex manifold with local complex coordinates $(T,X,Y,Z)$ and a complex non-holomorphic metric
\begin{align}
 g = \d{T}^2-\d{X}^2-\d{Y}^2-\d{Z}^2+\Phi(R,\bar{R})(L_{a}\d{x}^{a})^2, \label{CSK}
\end{align}
where $R=\sqrt{X^2+Y^2+Z^2}$ and 
\begin{align}
\Phi(R,\bar{R}) ={}& \frac{m}{R} + \frac{m}{\bar{R}} \label{KSf} \\
L_{a} \d x^{a} ={}& \frac{1}{\sqrt{2}}\left[ 
\d{T} + \left(\frac{1-|\zeta_{+}|^2}{1+|\zeta_{+}|^2} \right)\d{Z}
+ \left(\frac{\zeta_{+}+\bar{\zeta}_{+}}{1+|\zeta_{+}|^2} \right)\d{X}
+ \i\left(\frac{\zeta_{+}-\bar{\zeta}_{+}}{1+|\zeta_{+}|^2} \right)\d{Y}
\right], \\
\zeta_{+} ={}& \frac{-Z+R}{X+\i Y},
\end{align}
with $m$ a real parameter. Define two real slices 
$S_1 = \{T=t, \ X=x, \ Y=y, \ Z=z\}$,
$S_2 = \{T=t, \ X=x, \ Y=y, \ Z=z-\i a\}$, 
where $t,x,y,z$ are real and $a$ is a real parameter.
Then a calculation shows that $g$ restricted to $S_1$ 
is the Schwarzschild metric, and $g$ restricted to $S_2$ is 
the Kerr metric.

The above version of the trick singles out Schwarzschild and Kerr from a complex space by simply selecting two different real slices related in the usual Newman-Janis way $z\to z - \i a$ (in particular, it does not involve a change from spherical to spheroidal coordinates, see \cite{Visser}, and it also shows that the null vectors $L_{a}$ of Schwarzschild and Kerr correspond to two points of the Riemann sphere related by $z\to z - \i a$). However, the procedure is arbitrary in that, in going from Schwarzschild to Kerr, one still has to make the arbitrary replacement $\frac{2m}{r} \to \frac{m}{r}+\frac{m}{\bar{r}}$ in \eqref{KSf}.
Moreover, replacements of this sort, without any justification,
can be used to argue that any two metrics are ``related'' by a complex coordinate transformation. Let us illustrate this with a transformation from Schwarzschild to the C-metric:
\begin{proposition}
Consider a complex manifold with local complex coordinates $(U,V,W,\tilde{W})$ and a non-holomorphic metric
\begin{align}
g = \Omega^{-2}\left[ 2(\d{U}\d{V}+\d{W}\d\tilde{W})+\mathcal{F}(U,\bar{U})\d{V}^{2}+\mathcal{G}(W,\bar{W})\d\tilde{W}^2 \right], \label{SchwC}
\end{align}
where, separating $(U,V,W,\tilde{W})$ into real and imaginary parts according to $U=u+\i u'$, $V=v-\i v'$, $W=w+\i w'$, $\tilde{W}=\tilde{w}-\i\tilde{w}'$:
\begin{subequations}
\begin{align}
 \Omega(U,\bar{U},W,\bar{W}) ={}& u+\alpha(w'-u'), \\
 \mathcal{F}(U,\bar{U}) ={}& u^2(1-2mu) - (u'^2-1)(1+2m\alpha u'), \label{FDKS}\\
 \mathcal{G}(W,\bar{W}) ={}& 1-w^2 - (1-w'^2)(1+2\alpha m w'), \label{GDKS}
\end{align}
\end{subequations}
and $m, \alpha$ are two real parameters. Define two real slices by $S_{1}=\{U=u+\i, V=v, W = w+\i, \tilde{W}=\tilde{w}\}$ and $S_{2}=\{U=\i u', V=-\i v', W=1+\i w', \tilde{W}=-\i\tilde{w}'\}$. 
Then \eqref{SchwC} restricted to $S_{1}$ is the Schwarzschild metric, and \eqref{SchwC} restricted to $S_{2}$ is the C-metric.
\end{proposition}

To show this, notice first that
\begin{align}
g\big|_{S_{1}} ={}& \tfrac{1}{u^2} \left[ 2(\d{u}\d{v}+\d{w}\d\tilde{w})+u^{2}(1-2mu)\d{v}^2 + (1-w^2)\d\tilde{w}^2 \right], \label{Schw2} \\
g\big|_{S_{2}} ={}& \tfrac{1}{\alpha^2(w'-u')^{2}} \left[ 2(\d{u}'\d{v}'+\d{w}'\d\tilde{w}')+(u'^{2}-1)(1+\alpha m u')\d{v}'^2 + (1-w'^2)(1+2\alpha m w')\d\tilde{w}'^2 \right], \label{Cmet2}
\end{align}
and define new coordinates $(t_{\rm s},r,\theta,\phi)$ and $(\tau,x,y,\varphi)$ by 
\begin{align}
 & u =\frac{1}{r}, \qquad v = t_{\rm s}+\int\frac{\d{r}}{1-\frac{2m}{r}},
 \qquad w=-\cos\theta, \qquad 
 \tilde{w} = \i\phi - \int\frac{\d\theta}{\sin\theta}, \\ 
 & u' = -y, \qquad v' =\tau+\int\frac{\d{y}}{F(y)}, 
 \qquad 
 w' = x, \qquad \tilde{w}' = \i\varphi - \int\frac{\d{x}}{G(x)}
\end{align}
where $F(y)=(y^2-1)(1-2\alpha my)$, $G(x)=(1-x^2)(1+2\alpha m x)$. Then a short calculation gives the standard forms
\begin{align}
g\big|_{S_{1}} ={}& (1-\tfrac{2m}{r})\d{t}_{\rm s}^{2} - \frac{\d{r}^2}{(1-\frac{2m}{r})}-r^{2}(\d\theta^2+\sin\theta^2\d\phi^{2}), \label{Schw3} \\
g\big|_{S_{2}} ={}& \frac{1}{\alpha^2(x+y)^{2}} \left[ F(y)\d\tau^2 -\frac{\d{y}^2}{F(y)} - \frac{\d{x}^{2}}{G(x)} - G(x)\d\varphi^{2}\right]. \label{Cmet3}
\end{align}

The above shows that one can start from the Schwarzschild metric written in the form \eqref{Schw2}, ``complexify'' the functions $u^2(1-2mu)$, $(1-w^2)$ in such a way so as to obtain \eqref{FDKS} and \eqref{GDKS}, and then make a complex coordinate change (or choose a new real slice) and obtain the C-metric \eqref{Cmet2}, \eqref{Cmet3}. But this procedure is obviously completely arbitrary, since there are many ways of complexifying the functions in \eqref{Schw2}. 

We note that the (double-Kerr-Schild) form of the Schwarzschild and C metrics \eqref{Schw2}, \eqref{Cmet2} was found by using the facts that the space-times are conformally (Lorentzian) K\"ahler \cite{Flaherty} and that any K\"ahler metric has a double Kerr-Schild structure (as is not hard to show). 
The metrics inside the square brackets in \eqref{Schw2}, \eqref{Cmet2} are actually the K\"ahler metrics associated to these space-times \cite{AA}. 
(The fact that these solutions are double-Kerr-Schild is known from the work of Pleba\'nski and Demia\'nski \cite{PD}, but it is perhaps not straightforward to deduce \eqref{Schw2}-\eqref{Cmet2} from the expressions given in \cite{PD}.)

\section{Conclusions}
\label{Sec:Conclusions}

We gave a simple procedure for relating different solutions to the zero-rest-mass field equations via complex coordinate transformations, by exploiting the fact that the conformal group acts linearly on twistor space. In particular, we showed that a complex translation followed by a complex special conformal transformation of the (linearized) Schwarzschild field produces the (linearized) Pleba\'nski-Demia\'nski field. We also gave numerous other examples (constant fields, hopfions, waves, etc.) and (hopefully) provided a twistor intuition of why some of these transformations can be anticipated without calculation.

The fact that a complex translation of a point-like source produces a rotating source can already be intuitively anticipated from the Appell trick \cite{Appell} at the Newtonian level, while the Newman-Janis shift generalizes this to the relativistic level. Interestingly, a complex special conformal transformation has the effect of producing either {\em two} accelerating ring singularities, two accelerating point-like singularities, or a curious toroidal singularity (depending on the values of the parameters involved in the transformation). 
The apparent transformation of ``one object into two'' has to do with the fact that a special conformal transformation must be more properly applied in conformally compactified Minkowski space, and from the point of view of this space, a Coulomb field is actually double-valued (as it changes sign when crossing conformal infinity), see \cite[Section 9.4]{PR2}.

While our procedure for linear fields is unambiguous and essentially algorithmic, we argued that an analogous construction for non-linear fields is not so clear, at least not in the way in which the usual non-linear Newman-Janis shift is performed (that is, at the metric level). We illustrated this with a ``complex transformation'' that relates the (non-linear) Schwarzschild and C metrics, but we noticed that the procedure is completely artificial and non-unique. Part of the difficulty has to do with the fact that one is attempting to perform the complex transformation at the metric level, whereas in field theory the transformation is done in the curvature tensor (Maxwell, Weyl, and higher spin), which is a holomorphic object as it has definite chirality. In any case, since the recent applications of the Newman-Janis shift to amplitudes \cite{ArkaniHamed, Guevaraetal} make use of the field strength version of the trick, it is possible that the approach in this note can be applied to examine other kinds of  scattering processes.

\bigskip
\noindent
{\bf Acknowledgements.} 
It is a pleasure to thank Tim Adamo for very helpful comments on this manuscript and for conversations during a visit to the University of Edinburgh, in which parts of this work were first presented. I am also very grateful to the Alexander von Humboldt Foundation for support.

\appendix

\section{Conventions}
\label{Appendix}

The complexified spin group ${\rm SL}(2,\mathbb{C})\times{\rm SL}(2,\mathbb{C})$ has two independent basic representations, $(\frac{1}{2},0)$ and $(0,\frac{1}{2})$, we say that they have ``opposite'' chirality and we indicate this with the two different kinds of indices $A,B,...$ and $A',B',...$, which take values $0,1$ in both cases. The vector representation is $(\frac{1}{2},\frac{1}{2})\cong(\frac{1}{2},0)\otimes(0,\frac{1}{2})$, so it has indices $a\cong AA'$, $b\cong BB'$, etc. Accordingly, the (complexified)
Minkowskian coordinates $(t,x,y,z)$ of a point $x^{a}$
are encoded in the $2\times2$ matrix $x^{AA'}$ as
\begin{align}
x^{00'}=\tfrac{1}{\sqrt{2}}(t+z), \quad x^{01'}=\tfrac{1}{\sqrt{2}}(x+\i y), 
\quad
x^{10'}=\tfrac{1}{\sqrt{2}}(x-\i y), \quad x^{11'}=\tfrac{1}{\sqrt{2}}(t-z).
\end{align}
These components can be thought of as taken with respect to 
two constant spin dyads $(o_{A},\iota_{A})$, $(o_{A'},\iota_{A'})$ 
(which are in general not complex conjugates):
$x^{00'}=x^{AA'}o_{A}o_{A'}$, $x^{01'}=x^{AA'}o_{A}\iota_{A'}$, 
$x^{10'}=x^{AA'}\iota_{A}o_{A'}$, 
$x^{11'}=x^{AA'}\iota_{A}\iota_{A'}$. 
We raise and lower spinor indices with the (skew-symmetric) spin metrics $\epsilon_{AB}, \epsilon_{A'B'}$ and their inverses, according to $\omega^{A}=\epsilon^{AB}\omega_{B}$, $\varphi_{A}=\varphi^{B}\epsilon_{BA}$, etc.

Similarly, the Minkowskian components of
a vector field $V=V^{t}\partial_{t}+V^{x}\partial_x+V^{y}\partial_y 
+V^{z}\partial_{z}\cong(V^{t},V^{x},V^{y},V^{z})$ 
are equivalently encoded in the spinor components
\begin{align}
V^{00'}=\tfrac{1}{\sqrt{2}}(V^t+V^z), \quad 
V^{01'}=\tfrac{1}{\sqrt{2}}(V^x+\i V^y), \quad
V^{10'}=\tfrac{1}{\sqrt{2}}(V^x-\i V^y), \quad 
V^{11'}=\tfrac{1}{\sqrt{2}}(V^t-V^z).
\end{align}
The inverse transformation is
\begin{align}
V^t=\tfrac{(V^{00'}+V^{11'})}{\sqrt{2}}, \quad
V^x=\tfrac{(V^{01'}+V^{10'})}{\sqrt{2}}, \quad
V^y=\tfrac{(V^{01'}-V^{10'})}{\sqrt{2} \i}, \quad
V^z=\tfrac{(V^{00'}-V^{11'})}{\sqrt{2}}.
\end{align}
Putting $V_{AA'}=\epsilon_{AB}\epsilon_{A'B'}V^{BB'}$, we 
also have
\begin{align}
V_{00'}=V^{11'}, \quad V_{10'}=-V^{01'}, \quad 
V_{01'}=-V^{10'}, \quad V_{11'}=V^{00'}.
\end{align}

\end{document}